\documentclass[runningheads,a4paper]{llncs}
\usepackage{amssymb}
\usepackage{graphicx}
\usepackage{url}

\urldef{\mailsa}\path|{k.benning, m.menzel, j.reuter, m.axer}@fz-juelich.de|

\usepackage{textcomp}
\usepackage{amsmath}
\usepackage{upgreek}
\usepackage{marvosym}
\newcommand{\envelope}{\raisebox{4pt}{\scalebox{0.9}{\Letter}}\kern-1.7pt}
\usepackage{hyperref}
\usepackage{url}
\usepackage{float}
\usepackage{nicefrac}

\usepackage{booktabs}
\usepackage{tablefootnote}
\usepackage{calc}

\usepackage{microtype}

\usepackage{comment}

\usepackage{placeins}
\usepackage{float}
\usepackage{svg}

\usepackage{cancel}

\newcommand*{\horzbar}{\rule[.5ex]{2.5ex}{0.5pt}}

\usepackage{subfig}

\usepackage[
    backend=biber,
    style=numeric,
  ]{biblatex}
  
\begin{document}

\mainmatter  


\title{Independent Component Analysis\\for noise and artifact removal in\\Three-dimensional Polarized Light Imaging}


\titlerunning{ICA for noise and artifact removal in 3D-PLI}

%
%

\author{Kai Benning$^{1,2(}$\envelope$^)$ \and Miriam Menzel$^{1}$ \and Jan Reuter$^{1}$ \and Markus Axer$^{1,2}$}
\authorrunning{K.\ Benning et al.}


\institute{$^{1 }$Institute of Neuroscience and Medicine (INM-1), Forschungszentrum J{\"u}lich, \\52425 J{\"u}lich, Germany\\
\mailsa\\
$^{2 }$Department of Physics, Bergische Universit{\"a}t Wuppertal, \\42119 Wuppertal, Germany\\
}

%
%

\toctitle{ICA for noise and artifact removal in 3D-PLI}
\tocauthor{Kai Benning, Miriam Menzel, Jan Reuter, and Markus Axer}
\maketitle


\begin{abstract}
In recent years, Independent Component Analysis (ICA) has successfully been applied to remove noise and artifacts in images obtained from Three-dimensional Polarized Light Imaging (3D-PLI) at the meso-scale (i.e., $64$\,\textmu m). Here, we present an automatic denoising procedure for gray matter regions that allows to apply the ICA also to microscopic images, with reasonable computational effort. Apart from an automatic segmentation of gray matter regions, we applied the denoising procedure to several 3D-PLI images from a rat and a vervet monkey brain section.
\end{abstract}


\section{Introduction}

Studying the structure and function of the brain requires dedicated imaging techniques, allowing to map the highly complex nerve fiber architecture both with high resolution and over long distances. The neuroimaging technique \textit{Three-dimensional Polarized Light Imaging (3D-PLI)} \cite{axer2011a, axer2011b} was designed to reconstruct the three-dimensional orientations of nerve fibers in whole brain sections with micrometer resolution.

To remove noise and artifacts in 3D-PLI images, \textit{Independent Component Analysis (ICA)} has successfully been used \cite{Dammers2013, Dammers2013-book, Dammers2009}. However, the ICA has only been applied to mesoscopic images with a resolution of $64$\,\textmu m pixel size and not to microscopic images with a resolution of 1.33\,\textmu m pixel size so far. In order to resolve single nerve fibers, e.g. in the cerebral cortex, such a microscopic resolution is required. Light scattering, thermal effects, inhomogeneity of optical elements, or simply dust on the used filters  are noise sources, which combined with the weak birefringence 3D-PLI signal in cortical areas  inevitably lead to a low signal-to-noise ratio (SNR) and reconstruction errors.

Identifying and removing these noise components in microscopic 3D-PLI images is very challenging. The amount of data that has to be processed is extremely large and the sampling has to be done differently as compared to mesoscopic images. When applying the developed ICA method on microscopic images, the characteristic differences of the signal strengths in white matter and gray matter brain regions need to be taken into account. As the birefringence 3D-PLI signal of densly packed nerve fibers (i.e., fiber bundles) of the white matter proceeding within the sectioning plane are very strong and show a higher SNR than the less dense fiber tracts present in the gray matter, the denoising procedure needs only to be applied in regions of gray matter, which massively reduces the required computing time.

Here, we present an automatic ICA denoising procedure for gray matter areas in microscopic 3D-PLI images. It consists of an automatic segmentation of gray matter, followed by a data-parallel ICA artifact removal with automatic classification of noise and signal activations.


\section{Methods}

\subsection{Preparation of brain sections}

Brain sections from a Wistar rat (3 months old, male) and a vervet monkey (2.4 years old, male) were selected for evaluation.\footnote{All animal procedures have been approved by the institutional animal welfare committee at Forschungszentrum J{\"u}lich GmbH, Germany, and are in accordance with European Union guidelines for the use and care of laboratory animals. The vervet monkey brain was obtained when the animal was sacrificed to reduce the size of the colony, where it was maintained in accordance with the guidelines of the Directive 2010/63/eu of the European Parliament and of the Council on the protection of animals used for scientific purposes or the Wake Forest Institutional Animal Care and Use Committee IACUC \#A11-219. Euthanasia procedures conformed to the AVMA Guidelines for the Euthanasia of Animals.} 
The brains were removed from the skull within 24 hours after death, fixed in a buffered solution of 4\,\% formaldehyde for several weeks, cryo-protected with 2\,\% DMSO and a solution of 20\,\% glycerin, deeply frozen, and cut along the coronal plane into sections of 60\,\textmu m with a cryostat microtome (\textit{Polycut CM 3500, Leica, Microsystems}, Germany). The resulting brain sections were mounted onto a glass slide each, embedded in a solution of 20\,\% glycerin, cover-slipped, sealed with lacquer, and measured with 3D-PLI up to one day afterwards.

\subsection{Three-dimensional Polarized Light Imaging (3D-PLI)}

3D-PLI reconstructs the nerve fiber architecture of the brain with micrometer resolution. By transmitting linearly polarized light through unstained histological brain sections and analyzing the transmitted light with a circular analyzer, the birefringence of the brain section is measured, thus providing information about the three-dimensional orientations of the highly birefringent nerve fibers (myelinated axons) in the tissue \cite{axer2011a,axer2011b}.
The 3D-PLI measurements were performed with the same setup as described in \cite{menzel2020b} (\textit{LMP-1, Taorad GmbH}, Germany), using incoherent green light with a wavelength of about 550\,nm. During the measurement, the direction of polarization of the incoming light was rotated by $\rho = \{0^{\circ},10^{\circ}, \dots, 170^{\circ}\}$ and the transmitted light behind the circular analyzer was recorded by a CCD camera (\textit{Qimaging Retiga 4000R}) for each rotation angle, yielding a series of $N=18$ images.
The pixel size in object space was about 1.33\,\textmu m.
For each image pixel, the measured intensity values form a sinusoidal light intensity profile (\textit{PLI-signal}, $I(\rho)$).
The average value of the signal, i.\,e.\ the polarization-independent transmitted light intensity, is called \textit{transmittance} and is a measure for tissue absorption and scattering (highly scattering tissue components such as nerve fibers appear dark in the transmittance image). The amplitude of the normalized signal is called $retardation$ and indicates the strength of birefringence of the tissue. It is related to the out-of-plane angles of the nerve fibers in the brain section (in-plane nerve fibers show very high birefringence, while out-of-plane nerve fibers show much less \cite{menzel2015}). The phase of the signal indicates the in-plane \textit{direction} angle of the nerve fibers. Combining in-plane and out-of-plane angles, 3D-PLI allows to reconstruct the full three-dimensional orientations of the nerve fibers.


\subsection{Segmentation of white and gray matter}
\label{sec:masks}

Morphologically, brain tissue consists of two different tissue types: Gray matter and white matter. Gray matter contains various components, such as neuronal cell bodies, dendrites, synapses, glial cells, blood capillaries as well as myelinated and unmyelinated axons. Most of the gray matter regions are located at the outer surface of the brain (cortex), but also inner parts of the brain (i.e., sub-cortical nuclei) contain islands of gray matter. White matter is mainly composed of myelinated and unmyelinated axons. The largest portion of myelinated axons is located in the white matter.

For the ICA method presented here, it is necessary to consider gray matter regions separately from white matter regions.\footnote{Note that we here define white matter as all regions (/image pixels) that contain myelinated nerve fibers. Anatomically, some of these regions might be known as gray matter because they only contain a small amount of myelinated nerve fibers.}
In the following, we present a fully automated procedure to generate masks of white and gray matter.

\begin{figure}[htbp]
	\centering
	\includegraphics[width=3.5in]{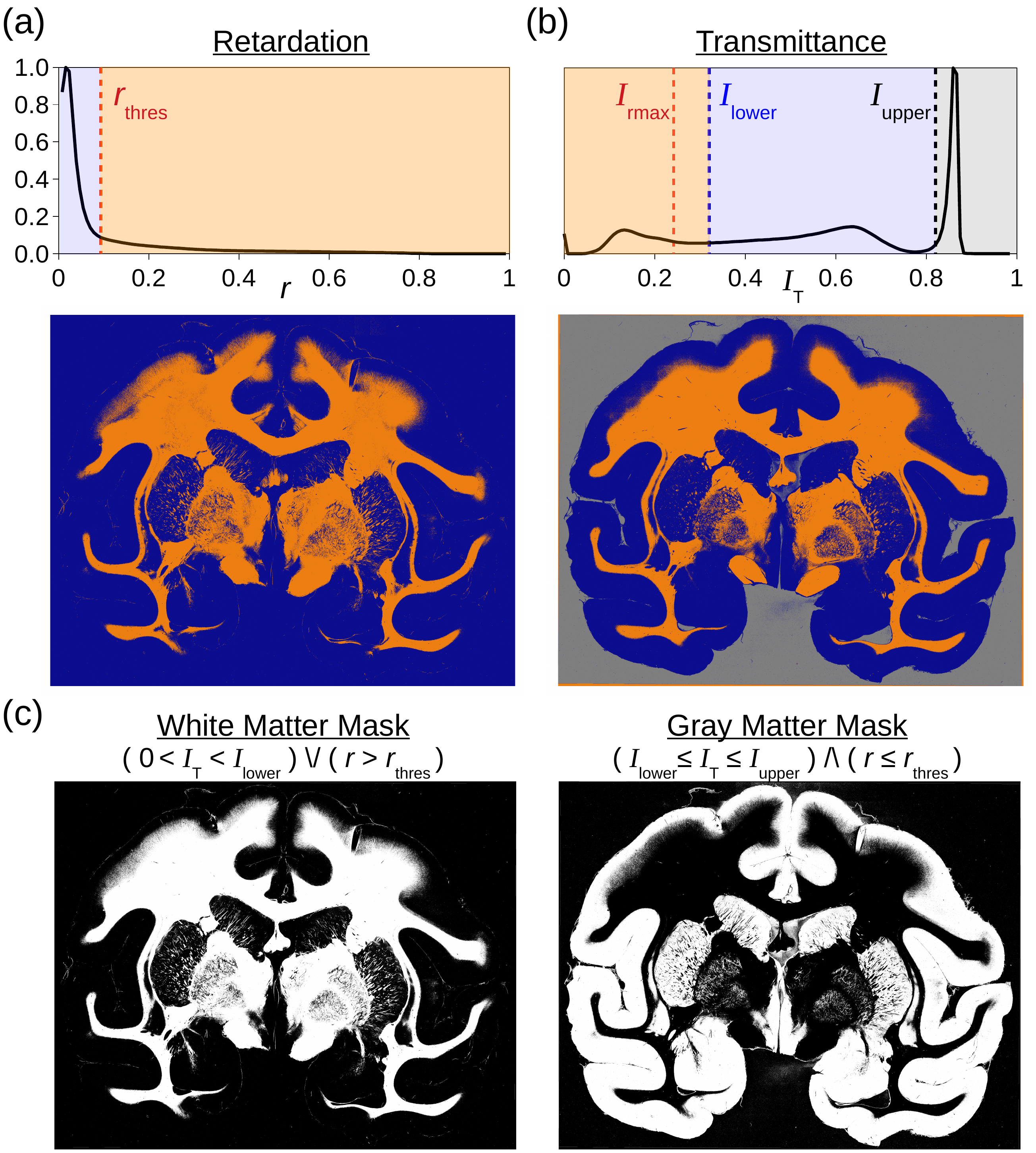}
	\caption{Mask generation for white and gray matter, shown exemplary for a coronal vervet monkey brain section. \textbf{(a,b)} Top: Histograms of retardation and transmittance images obtained from a 3D-PLI measurement (128 bins in [0,1]); the determined threshold values are marked by vertical dashed lines. Bottom: Image pixels with values belonging to the orange, blue, and gray shaded regions in the histograms are marked in the respective colors. \textbf{(c)} Masks for white and gray matter, computed using the threshold values defined in the histograms on top.}
	\label{fig:mask-generation}
\end{figure}

As nerve fibers (myelinated axons) are highly birefringent, all regions with high birefringence signals (i.\,e.\ large retardation values $r > r_{\text{thres}}$ in the 3D-PLI measurement, cf.\ \autoref{fig:mask-generation}(a) in orange) can be considered as white matter. (The determination of threshold values such as $r_{\text{thres}}$ will be described below.) On the other hand, regions with low birefringence signals ($r < r_{\text{thres}}$, \autoref{fig:mask-generation}(a) in blue) do not necessarily belong to gray matter, because regions with a small number of myelinated fibers, crossing nerve fibers, and nerve fibers that point out of the brain section (out-of-plane fibers) also yield low birefringence signals \cite{menzel2015}.

Studies by Menzel et al.\ \cite{menzel2020} have shown that regions with crossing nerve fibers and regions with in-plane parallel nerve fibers yield similar transmittance values $I_{\text{T}}$, while regions with out-of-plane nerve fibers show lower transmittance values. Gray matter regions, on the other hand, show notably higher transmittance values. Hence, we can use the transmittance value in the region with maximum retardation $I_{\text{rmax}}$ as a reference value (we expect that this region contains mostly in-plane parallel nerve fibers) and can then define all regions with similar or lower transmittance values as white matter ($0 < I_{\text{T}} \lesssim I_{\text{rmax}}$, containing crossing and out-of-plane fibers). All other brain regions are considered to be gray matter. To separate brain tissue from background, we make use of the fact that the transmittance in the recorded images is expected to be much higher outside of the tissue than within the tissue ($I_{\text{T}} > I_{\text{lower}}$, cf.\ \autoref{fig:mask-generation}(b) in gray).

To enable an automated segmentation into white and gray matter regions, we consider the retardation and transmittance histograms (consisting of 128 bins, see \autoref{fig:mask-generation} on top). Before computing the transmittance histogram, the values are normalized to $[0,1]$ and a median filter with circular kernel (radius of 10 pixels) is applied to the image to reduce noise. While the retardation histogram shows usually only a single peak at very low retardation values (caused by background and gray matter), the transmittance histogram shows one peak for low transmittance values (white matter), another peak for larger transmittance values (gray matter), and a third peak for high transmittance values (background).

To compute the threshold value $r_{\text{thres}}$ ($I_{\text{upper}}$), we determine the point of maximum curvature behind (before) the biggest peak in the retardation (transmittance) histogram, i.\,e.\ the position for which the angle difference between two neighboring data points becomes minimal. To ensure that the point of maximum curvature belongs to the onset of the biggest peak (and not to some other peak or outlier), we take the full width at half maximum (FWHM) of the peak into account and only search within $10\times$ FWHM behind the retardation peak and $2.5\times$ FWHM before the transmittance peak, taking the different forms of the histograms into account.

To compute $I_{\text{rmax}}$, the region with the maximum retardation value is determined. To ensure that the region belongs to a white matter region and is not an outlier caused by noise, we use the Connected Components algorithm from the OpenCV library \cite{bradski2000} (block-based binary algorithm using binary decision trees \cite{chang2015}) with eightfold connectivity.
We mark all pixels with maximum retardation value and count the number of pixels in the largest connected region. If the number is at least 512, we select this region as reference. If the number is lower, we reduce the maximum retardation value iteratively by 0.01, until we find such a region. In this reference region, we compute the average value in the normalized transmittance image ($I_{\text{rmax}}$, see red vertical line in \autoref{fig:mask-generation}(b)). This value can be used as first estimate to separate white from gray matter. To define the border more precisely, we use the point of maximum curvature between $I_{\text{rmax}}$ and $I_{\text{upper}}$ as new threshold value $I_{\text{lower}}$.

Taking all this into account, we can compute the masks for white and gray matter as follows:
\begin{align}
\text{White Matter:} \ \ \ &(0 < I_{\text{T}} < I_{\text{lower}}) \vee (r > r_{\text{thres}}), \label{eq:WM}\\
\text{Gray Matter:} \ \ \ &(I_{\text{lower}} \leq I_{\text{T}} \leq I_{\text{upper}}) \wedge (r \leq r_{\text{thres}}). \label{eq:GM}
\end{align}
All image pixels that fulfill \autoref{eq:WM} (\autoref{eq:GM}) are considered as white (gray) matter, see \autoref{fig:mask-generation}(b) in white. All other image pixels are considered as background.

\subsection{Independent Component Analysis (ICA)}

The ICA belongs to the group of \textit{Blind Source Separation (BSS)} techniques and can be used for data decomposition to find statistically independent components in a mixture of signals \cite{Hyvarinen2004}. ICA has been applied to various artifact removal tasks, e.\,g.\ ocular artifact removal in electroencephalography \cite{Dimigen2020}, cardiac artifact removal in magnetoencephalography \cite{Dammers2014}, and noise-signal-discrimination in functional magnetic resonance imaging \cite{McKeown2003}.

In 3D-PLI, a data set consists of a series of $N$ images (one for each rotation angle). To avoid that the background interferes with the decomposition, each image is divided into background and our region of interest (ROI) with the latter containing $M$ pixels. The measurements are flattened and centered to obtain a zero-mean data array $X$ with the dimension $N \times M$. The decomposition into sources $S$ with the shape $N \times M$ requires that the data can be represented as a linear mixture of independent signals without additional additive noise, that there exist sufficient samples for every extracted feature (general advise is to keep $k\cdot N^2 \geq M$ with $k \in \left\{ 1, 2, 3,\dots\right\}$), and that the distribution of the sources is non-Gaussian. With these prerequisites, the problem can be stated as

\begin{equation}
X = AS,
\end{equation}

where $A$ is the so-called mixing matrix with the dimension $N \times N$ and is yet unknown. Because $S$ and $A$ are both unknown, it is impossible to make a prediction about sign or amplitude of the basis vectors of $A$. Furthermore, we have no knowledge about the number of components in our data set, so we assume that the complexity of the data can be mapped by $N$ features.

Prior to performing the ICA, the data array $X$ is whitened by making use of a \textit{Principle Component Analysis (PCA)} \cite{Pearson1900} to lower the degrees of freedom to $N(N-1)/2$ \cite{Hyvarinen2004}. The ICA then estimates $W \approx A^{-1}$ by maximizing the entropy as in Infomax-based ICA \cite{Sejnowski1995} or by maximizing a measure of non-Gaussianity as in FastICA \cite{Hyvarinen1999,Hyvarinen2000}. We then obtain

\begin{equation}
    WX  = C \approx S,
\end{equation}

with the component vector $C$. We find the activation profiles of the Components in $W^{-1}$ as basis vectors. It was shown that 3D-PLI signals contain sub- as well as super-Gaussian independent components, therefore FastICA or Extended Infomax \cite{Lee1999}, an extension of the Infomax algorithm, can be used. This work uses the Extended-Infomax Implementation of the MNE-Toolbox \cite{mne-python}.

\subsection{Automatic noise removal with ICA}

\begin{figure}[H]
    \centering
    \includegraphics[width=3in]{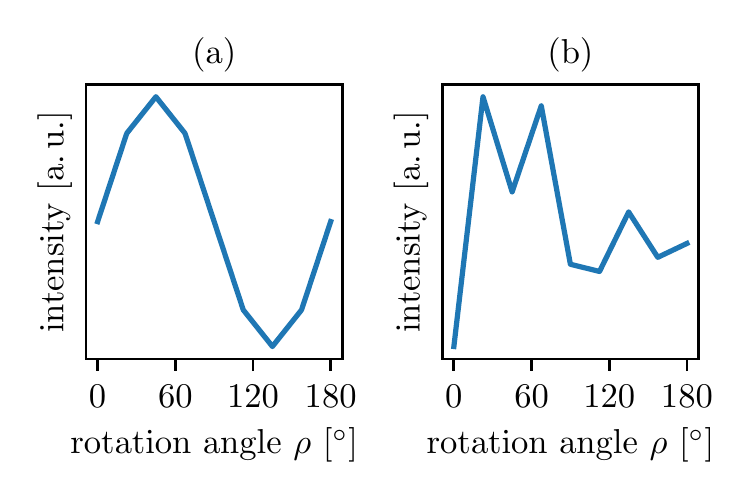}
    \caption{Archetypal signal discrimination for ICA artifact removal: \textbf{(a)} sinusoidal PLI-signal activation, \textbf{(b)} random PLI-noise activation from e.\,g.\ thermal noise or light scattering.}
    \label{fig:pli-discrimination}
\end{figure}

The activation profiles given by the ICA can be distinguished into two categories: noise activation profile and signal activation profile. Because we know the PLI-signal shape from theory, we know the shape of the basis vectors we are looking for in our mixing matrix $W^{-1}\approx A$. A simple classification problem is visualized in \autoref{fig:pli-discrimination}. The sinusoidal shaped activations are the ones to keep and we want to drop the activations that resemble random distributions.

The automatic identification is realized by fitting the expected (theoretical) function to each of the $N$ activations. As identification measure, the {mean squared error (MSE)} is calculated for every fit and compared to the mean of all MSE values. When the MSE of the $i$-th fit is smaller than $\nicefrac{1}{10}$ of the mean of all MSE values, we assume that the activation belongs to a signal component. Otherwise, we assume that the activation belongs to a noise component.

After the detection of all noise activations, we construct a denoised mixing matrix $W_{d}^{-1} \approx A_{d}$ by just zeroing out the respective column:

\begin{equation}
	W_{d}^{-1} = 
	\begin{pmatrix}
		\horzbar & \text{Signal Activation 1} & \horzbar \\
		\horzbar & \text{Signal Activation 2} & \horzbar \\
		         & \dots & \\ 
		\horzbar & \text{Signal Activation } J & \horzbar \\
		\horzbar & \cancel{\text{Noise Activation 1}} & \horzbar \\
		\horzbar & \cancel{\text{Noise Activation 2}} & \horzbar \\
		         & \dots & \\ 
		\horzbar & \cancel{\text{Noise Activation } K} & \horzbar
	\end{pmatrix}^{T} =
	\begin{pmatrix}
		\horzbar & \text{Signal Activation 1} & \horzbar \\
		\horzbar & \text{Signal Activation 2} & \horzbar\\
		         & \dots &  \\ 
		\horzbar & \text{Signal Activation } J & \horzbar \\
		\horzbar & \text{0} & \horzbar \\
		\horzbar & \text{0} & \horzbar \\
		         & \dots & \\ 
		\horzbar & \text{0} & \horzbar
	\end{pmatrix}^{T}.
\end{equation}

The denoised data array $X_d$ can then be obtained by remixing the previously unmixed components:

\begin{equation}
    W_{d}^{-1} C = W_{d}^{-1} W X = X_d.
\end{equation}

\subsubsection{Estimation of signal enhancement}

The noise reduction of the artifact removal is measured with a weighted chi-square statistic introduced by \cite{Dammers2013}. It is based on the reduced chi-squared statistic defined by

\begin{equation}
	\chi^{2} = \frac{1}{\nu} \sum_{i=1}^{N} \frac{(I(\rho_{i}) - f(\rho_i))^{2}}{\sigma(x,y,\rho)^{2}},
\end{equation}

with $\nu$ for the degrees of freedom, $N$ the number samples (here measurement angles), $I(\rho_i)$ the measured light intensity for an angle, $f(\rho_i)$ the expected function, and $\sigma(x,y,\rho)$ the standard error for every image point and angle. This statistic is applied to both noisy and denoised data leading to $\chi^{2}_{\text{raw}}$ and $\chi^{2}_{\text{ICA}}$. We denote the quotient of these two measures by \textit{relative Goodness of Fit (rGoF)}. In case one component is missing and the denoised signal is inherently different, an additional weighting factor $\omega$ defined by

\begin{equation}
	\omega = \frac{1}{\nu} \sum_{i=1}^{N} \frac{(f(\rho_{i}) - f^{*}(\rho_i))^{2}}{\sigma(x,y,\rho)^{2}}
\end{equation}

is included in the denominator, which penalizes large deviations between the expected function of the noisy signal $f(\rho_i)$ and the expected function of the denoised signal $f^{*}(\rho_i)$. The so obtained measure is denoted by \textit{weighted relative Goodness of Fit (wrGoF)}:

\begin{equation}
	\text{wrGoF} = \frac{\text{rGoF}}{\omega} = \frac{\chi^{2}_{\text{raw}}}{\chi^{2}_{\text{ICA}} \cdot \omega},
\end{equation}

where wrGoF $> 1$ is associated with a signal improvement while wrGoF $< 1$ is associated with a signal degradation.

\subsubsection{Parallelization concept}

The parallelization is implemented by distributing the workload of the ICA problem equally in $N$ parts to $N$ workers via the Message Passing Interface (MPI) with \texttt{mpi4py} \cite{mpi4py-1,mpi4py-2,mpi4py-3}. Every $n$-th element is sent to the $n$-th worker. After converging the result of all workers is collected and fused to the end result.

\section{Results}

\begin{figure}[h]
    \centering
    \subfloat[\centering Components $C_i$]{{\includegraphics[width=0.45\textwidth]{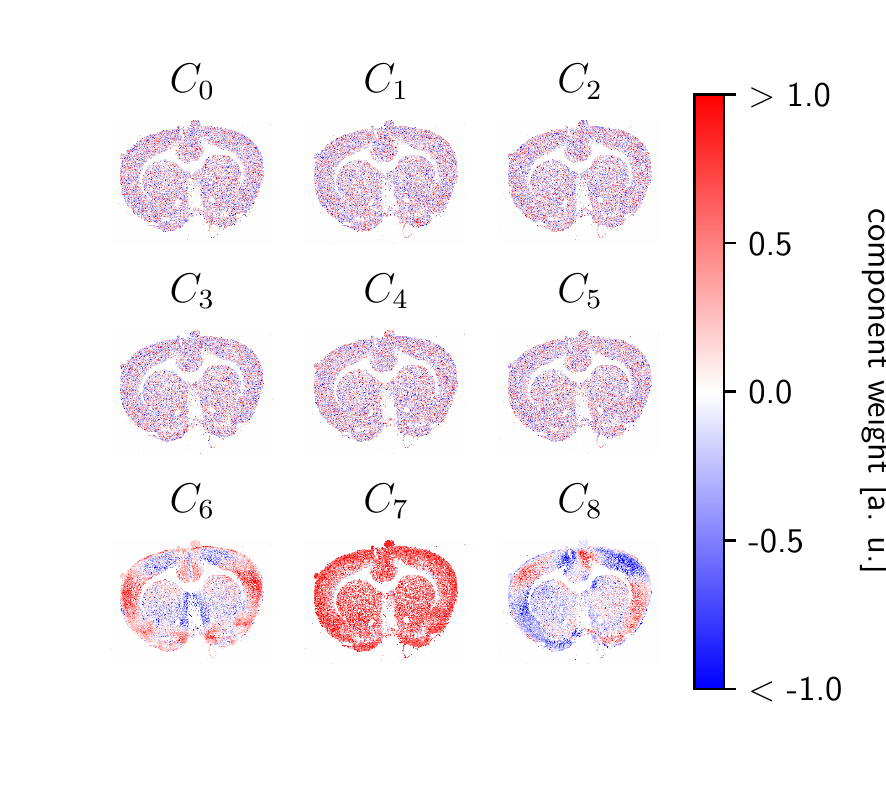} }}
    \subfloat[\centering Activations $A_i$]{{\includegraphics[width=0.45\textwidth]{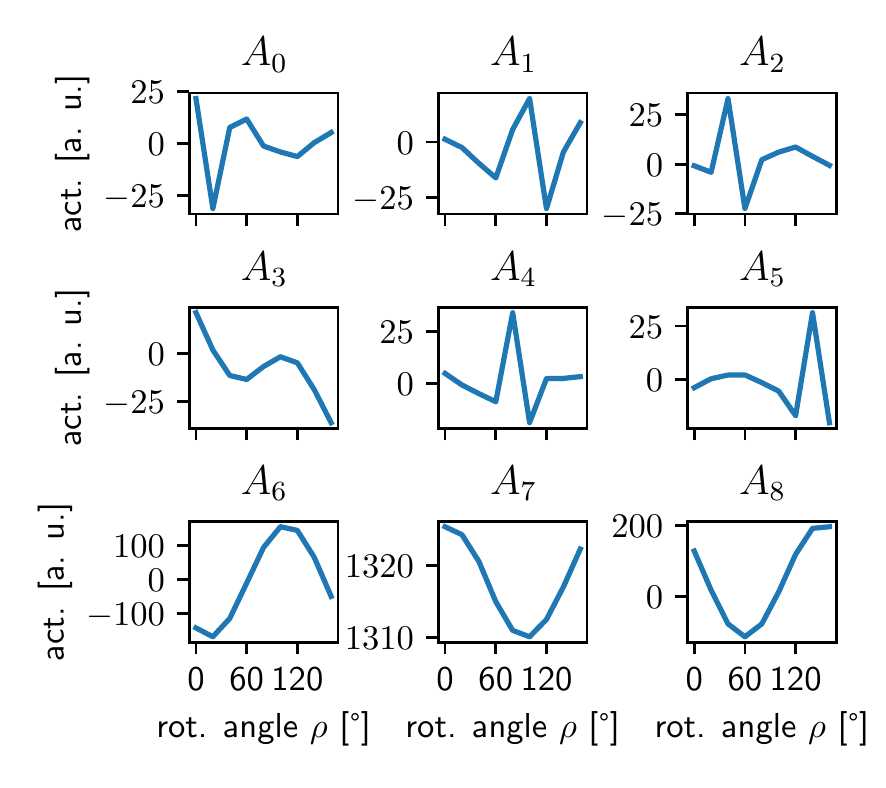} }}
    \caption{Independent Components $C_i$ for rat brain section no. 100 with their associated activations $A_i$. The noise components are in the first and second row, the signal components are in the last row.}
    \label{fig:components}
    
\end{figure}

The denoising procedure was applied to 22 brain sections (14 coronal rat brain sections and 9 coronal vervet brain sections). Every section was masked with a gray matter mask (as described in \autoref{sec:masks}) to remove background and white matter areas. In all cases, three components of interest were found, each with a sinusoidal activation function. The signal activations were kept and the noise activations were automatically removed. The resulting components and activations are shown exemplary for rat brain section no.\ 100 in \autoref{fig:components}.

\begin{figure}[h]
    \centering
    \includegraphics[width=0.85\textwidth]{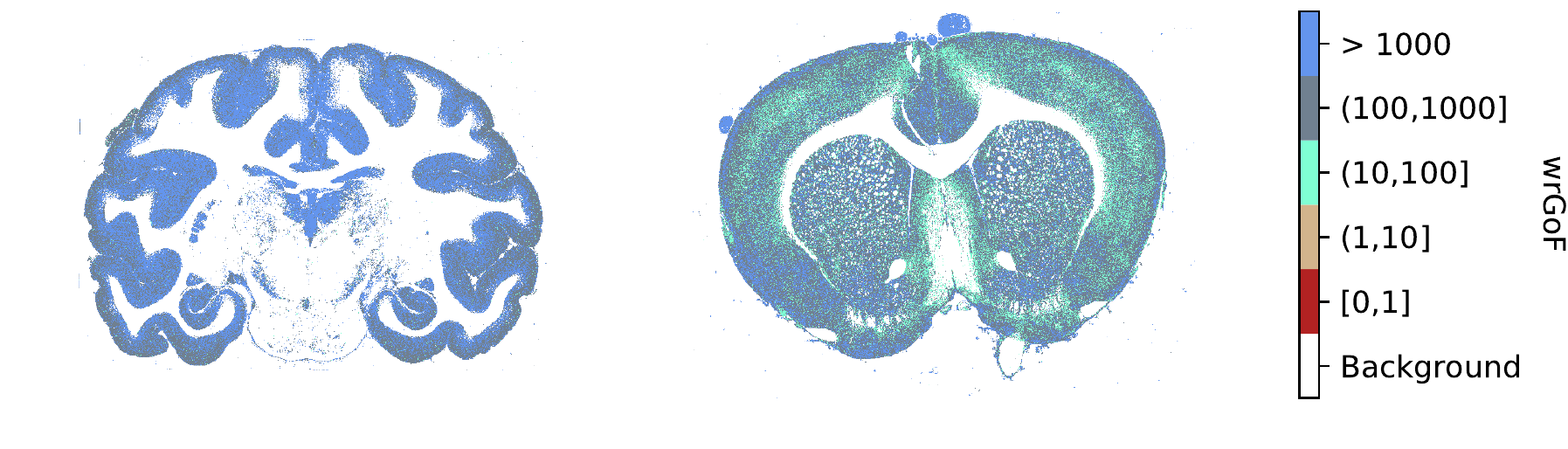}
    \caption{Signal enhancement in a brain section visualized via the wrGoF measure. \textbf{Left}: wrGoF-map of vervet brain section no.\ 627. \textbf{Right}: wrGoF-map of rat brain section no.\ 100.}
    \label{fig:wrgof}
\end{figure}

\begin{figure}[H]
    \centering
    \includegraphics[width=0.85\textwidth]{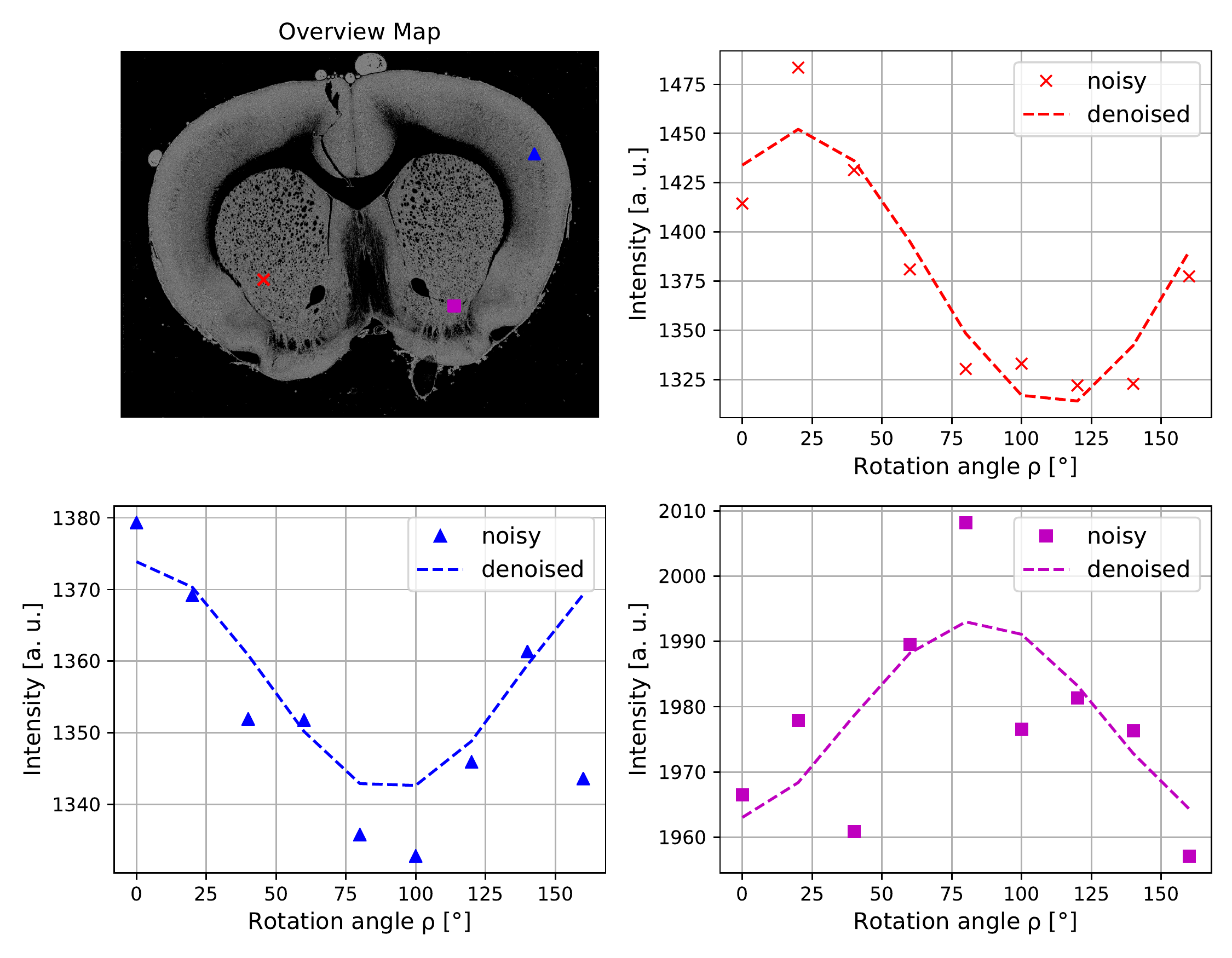}
    \caption{Signal enhancement by the ICA denoising procedure shown for rat brain section no.\ 100: The image on the upper left shows the transmittance of the brain section for gray matter (background and white matter are displayed in black). The graphs show the intensity profiles of three selected areas (colored dots in transmittance image) before and after the denoising procedure.}
    \label{fig:rat-improvment}
\end{figure}

The amount of wrGoF values are in all sections greater than one ($> 99.9\%$) and mostly greater than ten ($> 99\%$). A spatial distribution of the values is shown in \autoref{fig:wrgof} for the rat brain section (right) and a vervet brain section (left). In \autoref{fig:rat-improvment}, three selected intensity profiles are shown for the rat brain section. Each individual profile shows an improvement and the denoised profile describes the measurement in a smooth way and is not influenced by outliers. 

\begin{figure}[H]
    \centering
    \includegraphics[width=1.0\textwidth]{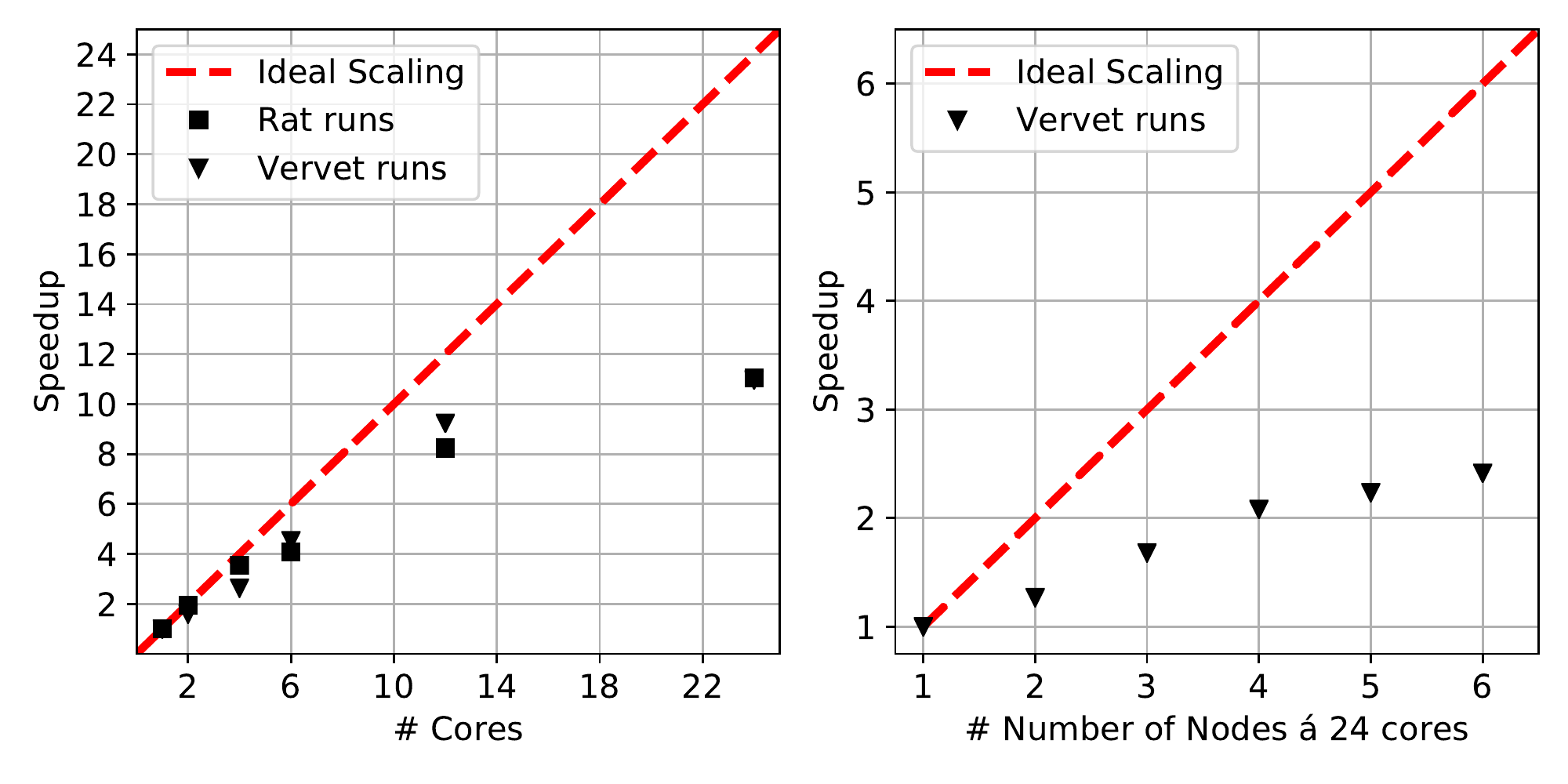}
    \caption{Scaling behavior of the ICA denoising procedure. \textbf{Left:} Intra-node scaling behavior for 1-24 Cores. \textbf{Right:} Global scaling behavior.}
    \label{fig:scaling-plot}
\end{figure}

The alternating parallelization approach achieves a linear speedup of up to 12 cores (i.\,e.\ half of the cores on a node on the JURECA supercomputer \cite{JURECA}). The usage of all cores on the node only improves the speedup by two additional units as seen in \autoref{fig:scaling-plot}. Overall, a weak scaling can be observed. While four nodes give a speedup of factor 2 with respect to one node, six nodes only offer a speedup of 2.5. For a complete vervet brain section (sample size $\sim 10^8$ pixels), the run time of the denoising routine for a single worker is about five hours. Using a whole node lowers this to half an hour, and using four nodes lowers the run time to 15 minutes.

This scaling behavior is the same for the rat and the vervet brain sections. Furthermore, the number of workers and therefore the number of partial ICA problems do not interfere with the quality of the denoising process. The percentage of wrGoF values greater than one ($> 99.9\%$) or greater than ten ($> 99\%$) are not influenced by the amount of parallelism.

\section{Discussion}

In this work, an automatic denoising procedure for 3D-PLI data based on Independent Component Analysis (ICA) for high-resolution PM data (with $1.33$\, \textmu m pixel size) was presented. Previous works studied the denoising of low-resolution LAP data (with $64$\, \textmu m pixel size) \cite{Dammers2013, Dammers2013-book, Dammers2009}, but the application on PM data was limited due to computational and memory constraints and was not fully automatic. Furthermore, the existing solutions were not suitable for a high-throughput workflow because masks for tissue had to be manually created or adjusted.

To overcome these limitations, three key steps had to been taken: The first step was to develop an automatic segmentation of brain tissue into white and gray matter, so that the ICA can work targeted on the noisy gray matter. The second step was an automatic detection of signal components in the ICA activations. The investigated brain sections showed good separability by a simple MSE measure. The zeroing of noisy components was straightforward to implement. Due to the fast convergence and easy separability, there was no need for constraints which would only complicate the procedure and add expensive hyperparameter training as in \cite{Dammers2013}. The third step was to parallelize the ICA in a pleasingly parallel manner to evenly distribute the workload and ensure that each worker receive similar statistics. This showed a weak, but significant scaling as shown in \autoref{fig:scaling-plot}.

The obtained results for the wrGoF measure were consistently better than the ICA denoising for LAP data presented in \cite{Dammers2013, Dammers2013-book}. The values were not influenced by the amount of parallelism. The amount of wrGoF values are in all brain sections greater than one ($> 99.9\%$) and mostly greater than ten ($> 99\%$). Overall, the results are very promising for high-throughput denoising of high-resolution 3D-PLI sections.

\subsubsection*{Acknowledgments.}

We thank Markus Cremer, Patrick Nysten, and Steffen Werner for the preparation of the histological brain sections. We also thank J\"urgen Dammers from the INM-4, Forschungszentrum J\"ulich, for introductory discussions about the specific application of ICA to 3D-PLI datasets.

Furthermore, we thank Karl Zilles and Roger Woods for collaboration in the vervet brain project (National Institutes of Health under Grant Agreement No.\ 4R01MH092311). 
This project has received funding from the European Union's Horizon 2020 Framework Programme for Research and Innovation under the Specific Grant Agreement No.\ 785907 and 945539 (``Human Brain Project'' SGA2 and SGA3).
We gratefully acknowledge the computing time granted through JARA-HPC on the supercomputer JURECA \cite{JURECA} at Forschungszentrum J\"ulich.

\printbibliography

\end{document}